\documentclass[ aip, amsmath,amssymb,reprint]{revtex4-1}

\usepackage{graphicx}
\usepackage{dcolumn}
\usepackage{multirow}
\usepackage{bm}
\usepackage{placeins}

\usepackage{amsmath}
\usepackage{graphicx}
\usepackage[colorinlistoftodos]{todonotes}
\usepackage[colorlinks=true, allcolors=blue]{hyperref}

\begin{document}
\title{ DFT study of undoped and As-doped Si nanowires approaching the bulk limit }
\author{Chathurangi Kumarasinghe}%
 \affiliation{London Centre for Nanotechnology, University College London, 17-19 Gordon St, London, WC1H 0AH}%
\email{c.kumarasinghe@ucl.ac.uk}
 
\author{David R. Bowler}%
 \affiliation{London Centre for Nanotechnology, University College London, 17-19 Gordon St, London, WC1H 0AH}%
 \affiliation{International Centre for Materials Nanoarchitectonics
   (MANA), National Institute for Materials Science (NIMS), 1-1
   Namiki, Tsukuba, Ibaraki 305-0044, Japan}
 \email{david.bowler@ucl.ac.uk}

\date{\today}

\begin{abstract}
The electronic properties of pure and As-doped Si nanowires (NWs) with radii
up to 9.53~nm are studied using large scale density functional theory
(DFT) calculations.  We show that, for the undoped NWs, the
DFT bandgap reduces with increasing diameter and converges to its bulk
value, a trend in agreement with experimental data.  Moreover, we show that
the atoms closest to the surface of the nanowire (NW) contribute less to
the states near the band edges, when compared with atoms close to the
centre; this is shown to be due to differences in Si-Si atomic
distances, as well as surface passivation effects.  When considering
As-doped Si NWs we show that dopant placement within the
NW plays an important
role in deciding electronic properties. We show that a low velocity
band is introduced by As doping, in the gap, but close to the
conduction band edge. The curvature of this low velocity band depends on the dopant location, with the curvature reducing when the dopant is placed closer to the center.  We also show that asymmetry of dopant location
with the NW leads to splitting of the valence band edge.
\end{abstract}

\maketitle

\section{Introduction}
Si NWs have electronic properties which are
different and far more attractive than bulk Si. More interestingly,
these properties could be engineered as required by controlling their
size, shape, composition and surface chemistry. High abundance of
Si(the second most abundant element on earth), non-toxicity and
availability of well-established fabrication techniques have further
increased interest in Si NWs for applications as diverse as photovoltaics
\cite{peng2011silicon,tsakalakos2007silicon},
thermoelectrics\cite{hochbaum2008enhanced,boukai2008silicon}, 
sensors \cite{miao2014silicon,cui2001nanowire,jee2014silicon} and
nanoelectronics \cite{cui2001functional,koo2005enhanced}. 

During the synthesis of Si NWs it is possible to engineer their
properties as required to suit different
applications\cite{cui2001diameter,morales1998laser,lew2004structural,cui2000doping,ma2001scanning}.
For instance, the bandgap of NWs is known to be sensitive to their
diameter, growth direction and the surface passivation method. 
Doping allows selective control of the Fermi level. Electronic band
structure that determines transport properties such as effective mass
and band overlap can be modified by NW size, shape and doping. Such
control is possible due to influence of high surface to volume ratio
and strong effects of quantum confinement seen as a particle size is
reduced.  

Doping is one of the essential steps to fabricate Si NW based
electronic devices. Boron(B) is a common p-type dopant
\cite{cui2000doping,cui2001nanowire} and phosphorus(P) and arsenic(As)
\cite{cui2001functional,colinge2010nanowire,tang2005situ} are common
n-type dopants used in Si NWs. As the diameter of Si NWs becomes
smaller, the positions of the dopants inside the NW as well as
the doping concentration, will likely influence the electronic
properties\cite{ng2011first}, but this needs to be investigated in
detail with atomistic methods.  

Theoretical analysis is necessary to provide  guidance in the
development and optimization of Si NWs for various
applications. Electronic properties of nanostuctures can be
theoretically studied with a reasonable balance between accuracy and
efficiency using DFT
based approaches. Due to the relatively high computational cost of
standard approaches, the majority of DFT simulations to date have been
limited to pure Si NWs with
diameters of around 5 nm or less
\cite{zhao2004quantum,ng2012chemically,zhuo2013surface,leu2008ab} and
doped Si NWs of even smaller diameters
\cite{peelaers2006formation,ng2012chemically,leao2008confinement,fernandez2006surface}. Since most
experimentally studied and practically used Si NWs are typically much
larger than this, the system sizes  theoretically studied need to be
significantly increased. With recent advances in computer power
and DFT methods, it is possible to reasonably accurately model such
larger scale systems. This allows us to study properties of Si NWs with
practical dimensions (up to 10~nm)\cite{ORourke:2018um}.

In this work we study the atomic and electronic structure of Si NWs
with diameters close to experimental NWs, using the large scale DFT code
\textsc{Conquest}\cite{bowler2002recent,bowler2010calculations}, which enables
modelling of systems with many thousands of atoms without approximation. We study the
electronic properties of H-passivated Si NWs
grown in the [110] direction, both undoped and doped with As. This particular growth direction is
chosen because it is known from experimental observations that [110]
crystallographic orientation is the preferred direction of wire
growth for diameters smaller than
10~nm\cite{ma2003small,wu2004controlled}. Most current experimental
techniques produce Si NWs with surface dangling bonds passivated either
oxides\cite{wang1999si,cui2001diameter} or with hydrogen
\cite{wu2004controlled,ma2003small}.  As the creation of a realistic
oxide layer (and sampling of different possible amorphous
configurations to adequately represent possible structures) would
generate an entirely new research project, we choose to focus on the
simplest possible termination: hydrogen passivated Si NWs. 

We are particularly interested in the electronic 
structure of pure and As-doped NWs, as affected by the diameter
of the NWs and the location of the dopants.
The smallest of our undoped NW models is approximately 2.61~nm in
diameter, containing 612 atoms, and the largest is approximately
10.2~nm in diameter and contains 2172 atoms, in the computational simulation
cell (axial repeat length of 4 Si layers).
To the best of our knowledge, the largest undoped Si NW studied
in literature with first-principles techniques has a diameter of only
to 7.3~nm \cite{ng2011first}. The diameter of NWs are usually
calculated as the diameter of the circle that can contain all Si atoms
of the NW (excluding the H atoms in the surface). 

Arsenic(As) is a n-type dopant that can enter the silicon lattice
substitutionally and it is commonly used to dope Si
NWs\cite{sivakov2010roughness,colinge2010nanowire}. It is known
to diffuse less in the bulk Si than other dopants such as B,P or Al
\cite{selberherr2012analysis}, therefore it can be useful in
situations where dopant stability is crucial. However, unlike B or P
doped cases, very few theoretical investigations have been conducted
to date for As doped Si NWs. The largest  As-doped Si NW 
in previous first-principles studies has a diameter $<2$~nm
\cite{reenu2015dft}, which is far from experimental samples. Here we
study As-doped  Si NWs with 
diameters ranging from approximately 2.61~nm to 7.34~nm, resulting in
612 to 2,472 atoms respectively, in a computational simulation cell with an
axial repeat length of 8 Si layers.

\section{Methods}

\begin{figure}
\centering
\includegraphics[width=0.45\textwidth]{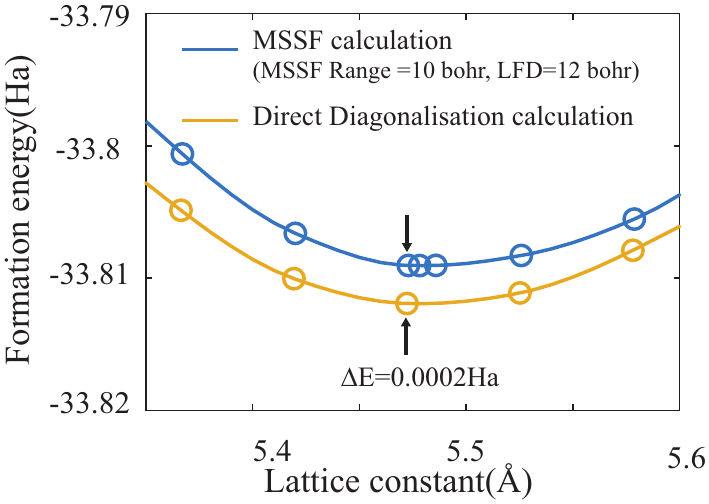}
\caption{\label{fig:EOS} Variation of energy with bulk Si lattice
  constant calculated using multi-site support function method
  (blue/dark lines) and direct diagonalisation method (beige/light
  lines).}  
\end{figure}

\begin{figure*}
\centering
\includegraphics[width=1\textwidth]{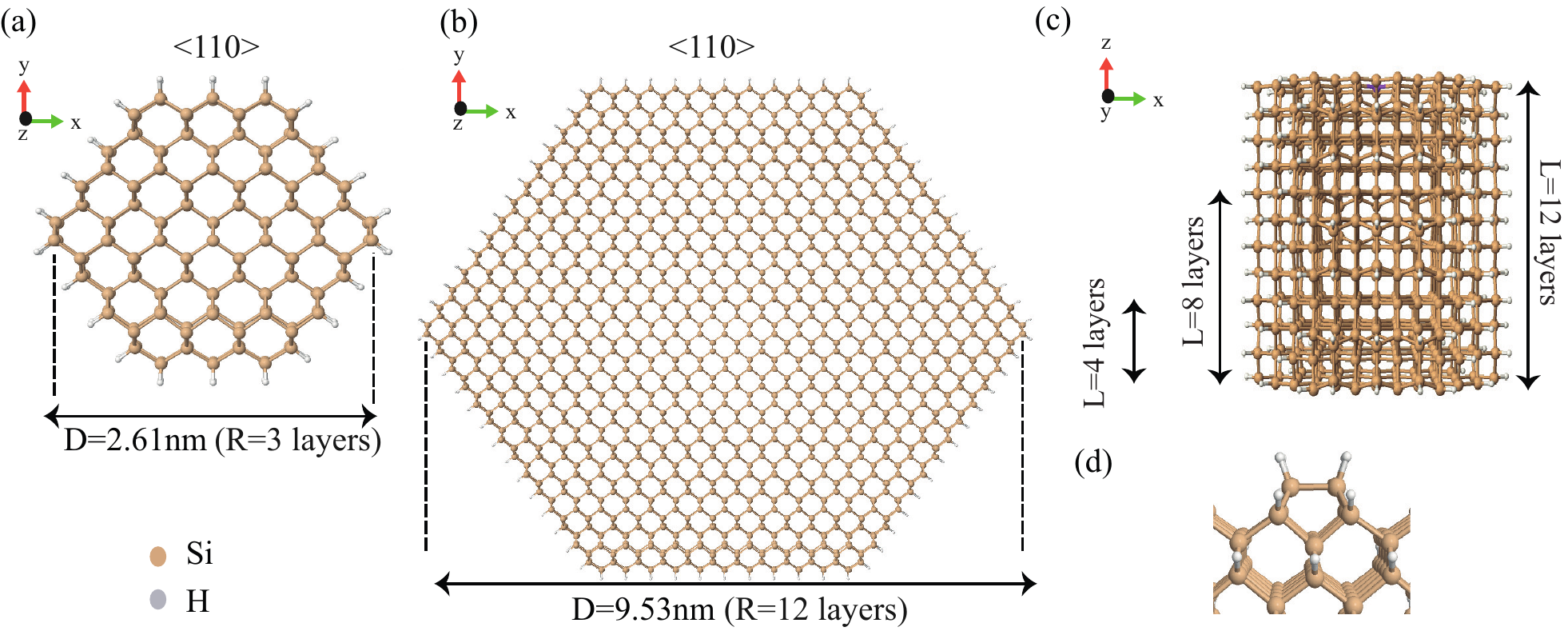}
\caption{\label{fig:Structure} Atomic structure of [110] oriented, H passivated
  pure silicon NWs. The approximate diameter in nanometers is
  indicated by D; the radius (counted as the number of atomic layers
  from the outermost Si layer to the center of the NW) is indicated by
  R; and the longitudinal length of the simulation cell (counted as the
  number of atomic layers) is indicated by L. (a) and (b) 
  cross-sectional views of smallest($\approx$ 2.61~nm) and the
  largest($\approx$ 9.53~nm) diameter NWs used, respectively. (c)
  Axial view of the smallest NW, showing the three different lengths
  used. (d) detail of the passivated surface reconstruction of the
  (100) surface of the NWs.}  
\end{figure*}

We have performed DFT calculations for the pure and doped Si NWs with
\textsc{Conquest}\cite{bowler2010calculations,
  bowler2012methods,bowler2002recent} using multi-site support
functions\cite{nakata2014efficient}. Since the details of the
implementation have been discussed in detail elsewhere, we summarize
only the main principles needed to explain the current approach. 

\textsc{Conquest} is a large-scale DFT code, which uses a
local-orbital basis\cite{Torralba:2008wm}.  While the code is capable
of linear scaling operation\cite{bowler2012methods}, we here perform
exact diagonalisation, using the multi-site support function
approach\cite{nakata2014efficient,Nakata2015} to allow simulation of
several thousand atoms on relatively modest computational resources.

The Kohn-Sham wavefunctions are represented by support functions,
$\phi_{i\alpha}(\mathbf{r})$, which are also used to form the
Hamiltonian and overlap matrices.  The support functions themselves
are represented in terms of basis functions, in this case
pseudo-atomic orbitals (PAOs):
\begin{equation}
\phi_{i\alpha}(r)=\sum_{\mu} c_{i\alpha,i\mu} \chi_{i\mu}(r),
\end{equation}
where $\chi_{i\mu}(r)$ is a PAO on atom $i$, with a composite index
$\mu$. The accuracy of the calculations can be improved by increasing
the number of PAOs on each atom (though it is hard to do this in a
systematic manner); however, this increases matrix sizes and
makes the diagonalisation computationally expensive.

As an alternative, the accuracy can be increased even with a
small number of support functions, by using  multi-site support
functions. Multi-site support functions are constructed as linear
combinations of PAOs of a target atom and its neighbouring atoms and
can be written as, 
\begin{equation}
\phi_{i\alpha}(r)=\sum_{k\in r_{MS}}\sum_{\mu} C_{i\alpha,k\mu} \chi_{k\mu}(r).
\end{equation}
The summation over atoms $k$ include the target atom $i$ and its
neighbouring atoms within a radius $r_{MS}$. The coefficients
$C_{i\alpha,k\mu}$ are found by using 
a localized filter diagonalization method
\cite{rayson2009highly,rayson2010rapid}. This local diagonalisation
included all atoms in a range $r_{LD}>=r_{MS}$, but only
the coefficients from atoms $r_{MS}$ range are used as
$C_{i\alpha,k\mu}$. Using a larger $r_{MS}$ and $r_{LD}$ improves the
accuracy of the calculations but increases the computational costs.  

In finding the DFT ground state, the coefficients $C_{i\alpha,k\mu}$
are calculated for a given density and self-consistent-field
calculations are performed for these coefficients. Once self
consistency is reached, the coefficients are updated. This process is
repeated until the energy is converged to within a certain threshold,
reaching the ground state. We note that, while \textsc{Conquest} is
capable of calculations with linear scaling cost, here we have used
exact diagonalization as the system is relatively small. 

We used a large basis sets for all elements, with three radial
functions (also known as zeta functions) for all angular momenta,
including valence electrons and a polarisation shell (triple zeta,
triple polarisation, or TZTP).  For both silicon and arsenic this is
equivalent to 27 PAOs consisting of three s, three p, and
three d orbitals, which were contracted down to 4 multisite support
functions.  For hydrogen, the basis consisted of 12 PAOs (three s and
three p) contracted down to one multisite support function.  We
project the density of states (DOS) onto atoms using the basis
function coefficient contribution to each band.

To identify optimum values for $r_{MS}$ and $r_{LD}$, we
investigated the energy convergence of bulk silicon with respect to
these parameters.  We found that $r_{MS}$ =10\,$a_0$ and $r_{LD}$
=12\,$a_0$ gave an energy convergence within an order of
$10^{-3}$~eV, and used these parameters for all elements.

As a further test of the multi-site radii, and to characterise the
calculations, the equation of state for bulk silicon was calculated
using multisite support functions, and compared to diagonalisation in
the primitive PAO basis set, as shown in Fig.~\ref{fig:EOS}.
The difference in energy at the minimum is less than $2\times 10^{-4}$
Ha ($\sim 5\times 10^{-3}$\,eV), while the lattice constants have the
same value, 5.47\AA. 
This confirms the accuracy of
our multi-site support function parameters.
We used this lattice constant for the NWs in the axial direction.

The generalized gradient approximation
(GGA) in the form of the PBE exchange-correlation functional
\cite{perdew1996generalized} was used in all calculations.
We note that this choice of functional will result in inaccuracies in
the band gap calculated (as seen below in
Sec.~\ref{sec:pure-si-nws}).  Use of a hybrid functional would
improve the band gap, though a screened functional might be necessary
to account for the extension of the system\cite{Janesko:2009jb}, and
it is not always clear what fraction of exchange is needed in a given system.
While there might be some changes to the curvature of the bands, the
largest effect on the band structure would be to open the gap.
While most
integrals are performed analytically or radially, we still require a real-space
grid for some matrix elements, and we set the 
integration grid cutoff to 80~Ha, which was identified as sufficient
for the DFT total energy to converge within $10^{-5}$~eV. Structural
relaxations used a Monkhorst-Pack grid with a density of $1\times
1\times 2$, where the NW is assumed to lie along the z-axis.  The NW
was constrained along the [110] axial direction, but is free to expand
along the radial direction. The Si NW is put in a unit cell with more
than 13~\AA\ vacuum spacing in the transverse directions to avoid any
interactions between the neighboring image NWs due to
periodicity. The Si NWs are relaxed until the absolute value of force
acting any atom is less than 0.02 eV/\AA\ using the FIRE algorithm
\cite{bitzek2006structural}.

\section{Results and Discussion}

\subsection{Pure Si NWs}
\label{sec:pure-si-nws}

\begin{figure}
\centering
\includegraphics[width=0.45\textwidth]{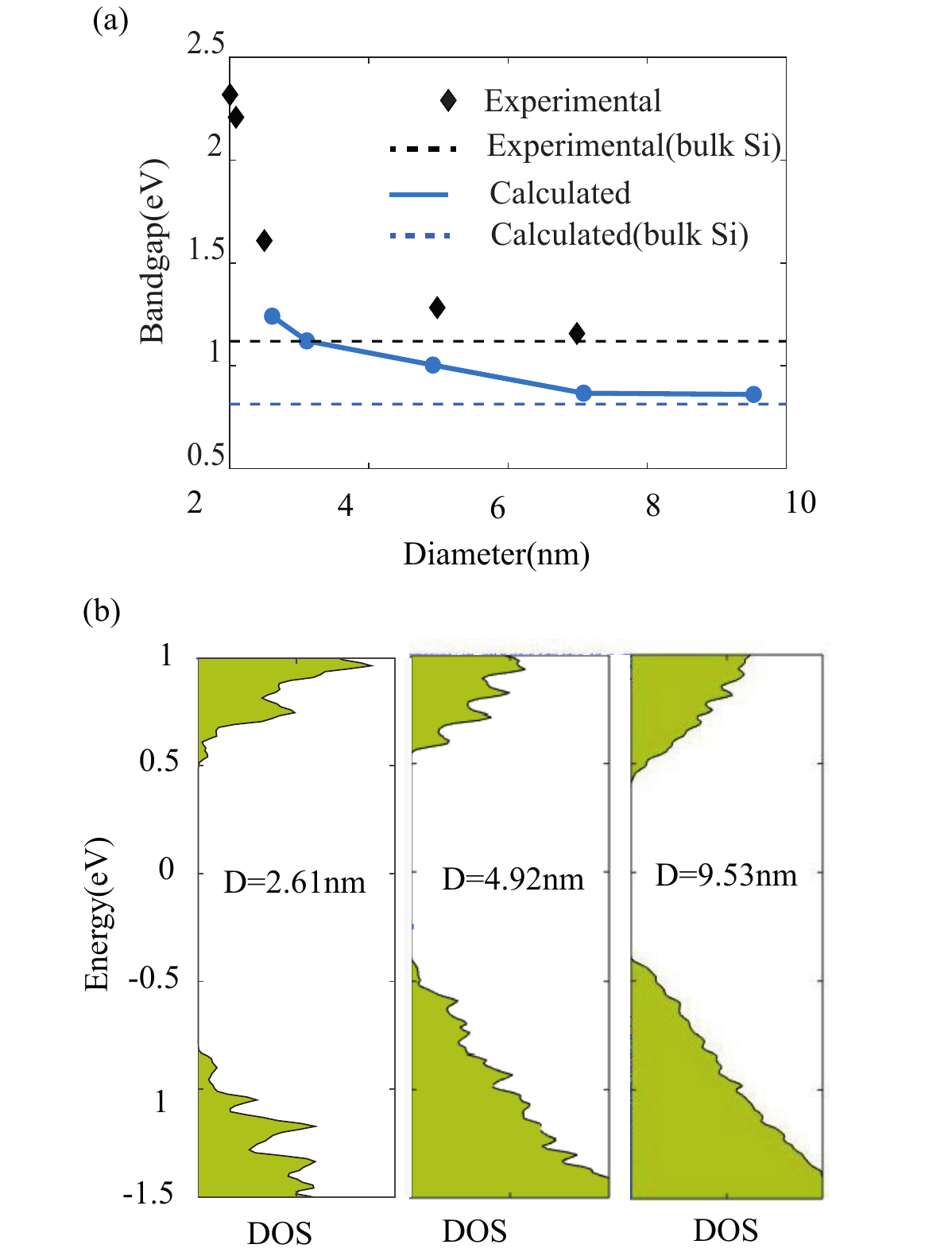}
\caption{\label{fig:gap} (a)Band gap as a function of diameter
  showing experimental measurements of small NWs\cite{ma2003small}
  (dark diamonds), our calculated gaps (light circles) and the bulk
  silicon gap, both experimental and calculated. (b)
  Density of states(DOS) of pure Si NWs with diameters of 2.61~nm,
  4.92~nm and 9.53~nm.} 
\end{figure}

\begin{figure*}
\centering
\includegraphics[width=0.95\textwidth]{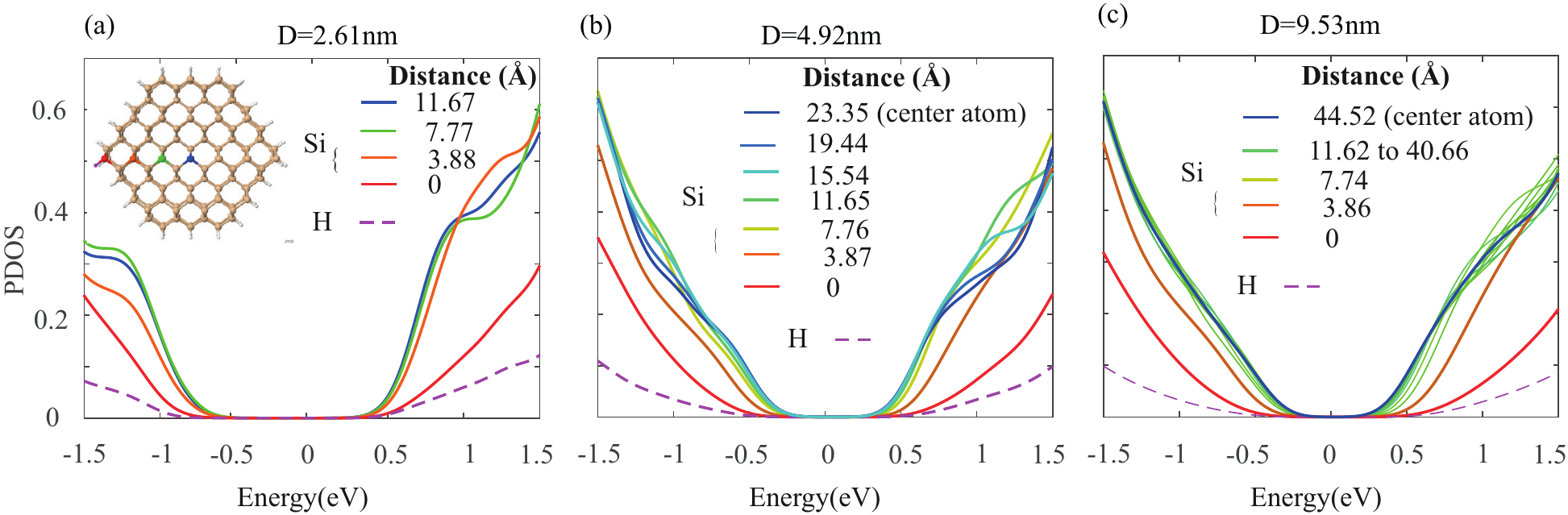}
\caption{\label{fig:SiNWPDOS} Partial density of states (PDOS) for
  Si atoms at different positions in NWs of different
  diameters. Distances (in \AA ) are given relative to the
  outermost Si layer the NW, as illustrated in (a). PDOS of Si atoms
  are represented by solid lines while PDOS from surface H atoms are
  represented by dashed lines. (a),(b) and (c) show PDOS for silicon
  atoms in NWs with
  diameters of 2.61~nm, 4.92~nm and 9.53~nm,
  respectively.} 
\end{figure*}

The most frequently reported growth directions for Si NWs are [110] and
[111] directions, and less frequently, the [112]
direction\cite{wu2004controlled,ma2003small}. Si NWs with diameters
in the range of our study primarily grow along the [110]
direction\cite{wu2004controlled,ma2003small}, with a hexagonal cross
section, bounded by four (111) and two (001) surface
facets. Larger diameter NWs are known to grow in [112] and [111]
directions\cite{wu2004controlled}, but we will not consider them
here.

Cross-sectional models of the smallest and largest diameter pure Si
NWs considered in this work are shown in Fig.~\ref{fig:Structure}(a) and
~\ref{fig:Structure}(b) respectively. The dangling bonds of Si atoms on the
surface are passivated by H, as illustrated in Fig.~\ref{fig:Structure}(d); we
note that the (001) surface is dimerised, with a monohydride
termination. An axial view of the supercell used for the NWs is
shown in Fig.~\ref{fig:Structure}(c), illustrating the three lengths
considered: one (L=4 layers), two (L=8 layers) and three primitive cells
(L=12 layers) along the axial direction.

We start by considering the effect of the NW diameter on the
electronic structure of pure silicon; for this part of the study, we
use only one repeat length along the axis.  Figure~\ref{fig:gap}(a) shows the
band gap of pure Si NWs of different diameters, ranging from
2.61~nm to 9.53~nm. The sharp increase in the band gap with the
reduction of diameter can be attributed to quantum confinement
effects. Our calculated values follows the trend seen in the
experimental data\cite{ma2003small} very well.  For bulk silicon, the
experimental band gap at room temperature is 1.12~eV. Experimentally
obtained bandgap values converge to this bulk value with increasing
diameter. Similar behaviour can be seen in our calculated values,
though these are affected by the well-known problem with band gaps in
DFT.  Our NW gap does converge to the calculated bulk band gap
with increasing diameter.

Going into more detail, we also see that the density of states for the full
NW is strongly affected by the NW diameter in
Fig.~\ref{fig:gap}(b). The smallest diameter NW has a very high
variation in electronic states with changing energy, while this
variation becomes smoother with the higher diameter NWs (this is in part due to
the larger number of atoms included in modelling).

In charge transport related applications, the states near the edges of
conduction and valence bands dominate the transport
properties. Therefore, it is important to understand which atoms
contribute to the edges of the conduction valence bands
most. Figure~\ref{fig:SiNWPDOS} shows partial densities of states projected
onto specific atoms at different distances from the surface of the
NWs, concentrating on the region around the Fermi level. Due to
the difference in strain and surrounding environment experienced by
the atoms based on their position, the contribution is not the
same. It can be seen that, regardless of the diameter, Si atoms in the
two layers closest to the surface contribute least to the band
edges, while the atoms at the center of the NW have a higher
contribution. When considering the contributions within 0.5~eV of the
band edges, it can be noted that the magnitude of the contribution is
inversely proportional to the distance from the surface. Moreover, the
H atoms used for passivating the dangling bonds have a negligible
contribution, when compared with the Si atoms (this is entirely
expected if they function purely as passivation).  Electronic
transport will therefore likely be largely far from the surface of the
Si NWs. The Si-Si distance 
near the surface is slightly more than the Si-Si distance near the
centre. The influence from the H atoms in combination with this factor
leads to the PDOS behaviour observed. Similar PDOS characteristics
have been seen in other NW studies\cite{ng2012chemically}.

\subsection{As-doped Si NWs}

We now turn to doping; we study As-doped Si NWs with radii of 3 ,6 and 9 layers
(D=2.61, 4.92 and 7.34~nm) 
with the aim of understanding the the role played by the
dopants and its positioning on the electronic band structure. These
positions are selected with the intention of identifying different
effects when the dopants are close to the surface and away from the
surface.  We limit the radii compared to the pristine NWs (which
extended to 12 layers) so that we can extend the axial size of the
NW to isolate the dopants.  Our tests showed that the change
in radius from 9 to 12 layers gives only very small changes in
electronic structure (e.g. the gap in Fig.~\ref{fig:gap}).

\begin{figure*}
\centering
\includegraphics[width=0.95\textwidth]{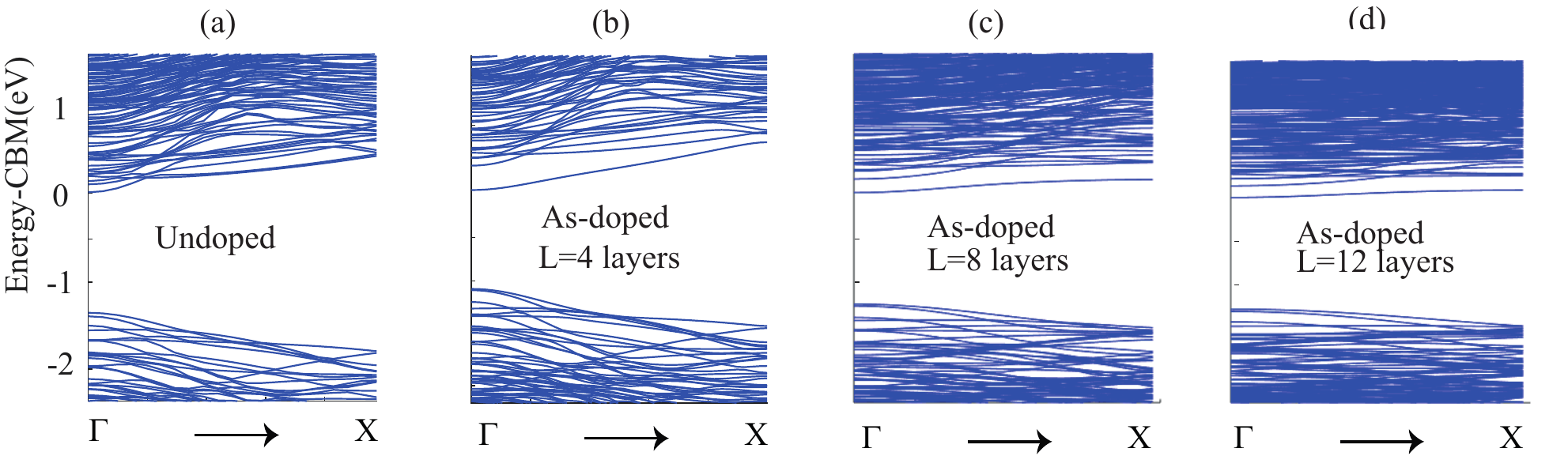}
\caption{\label{fig:bands} Comparison  of band structure along the axial
  direction of an undoped (a) and doped (b-d) Si NWs with different
  doping densities, and diameter R=3
  layers.  The As dopant is in the
  center (shown in Fig.~\ref{fig:PDOSR3L8}(a),As position 3).(b),(c) and (d)
  represent NWs with  L=4, 8 
  and 12 atomic layers, respectively. The band energies are all given
  relative to the respective conduction band minima.} 
\end{figure*}

We first test the effect of increasing the \emph{axial} size of the
NW, thus probing the isolation of the dopant along the axis.
Figure ~\ref{fig:bands} compares the bandstructures of an undoped NW 
with doped NWs of different axial lengths (and hence different As
doping concentrations). Different doping 
concentrations were implemented using 1 repeat length (L=4 layers, 0.77\,nm),
2 repeat lengths (L=8layers, 1.55\,nm) and
3 repeat lengths (L=12 layers, 2.32\,nm) in the axial direction of the supercell. The
supercell contains a single As atom. The As to Si ratio in each case
are 1/159, 1/319 and 1/479 respectively. We have used our smallest
diameter NW model which has a radius (R) of 3 layers for these
calculations. 

\begin{figure}
\centering
\includegraphics[width=0.48\textwidth]{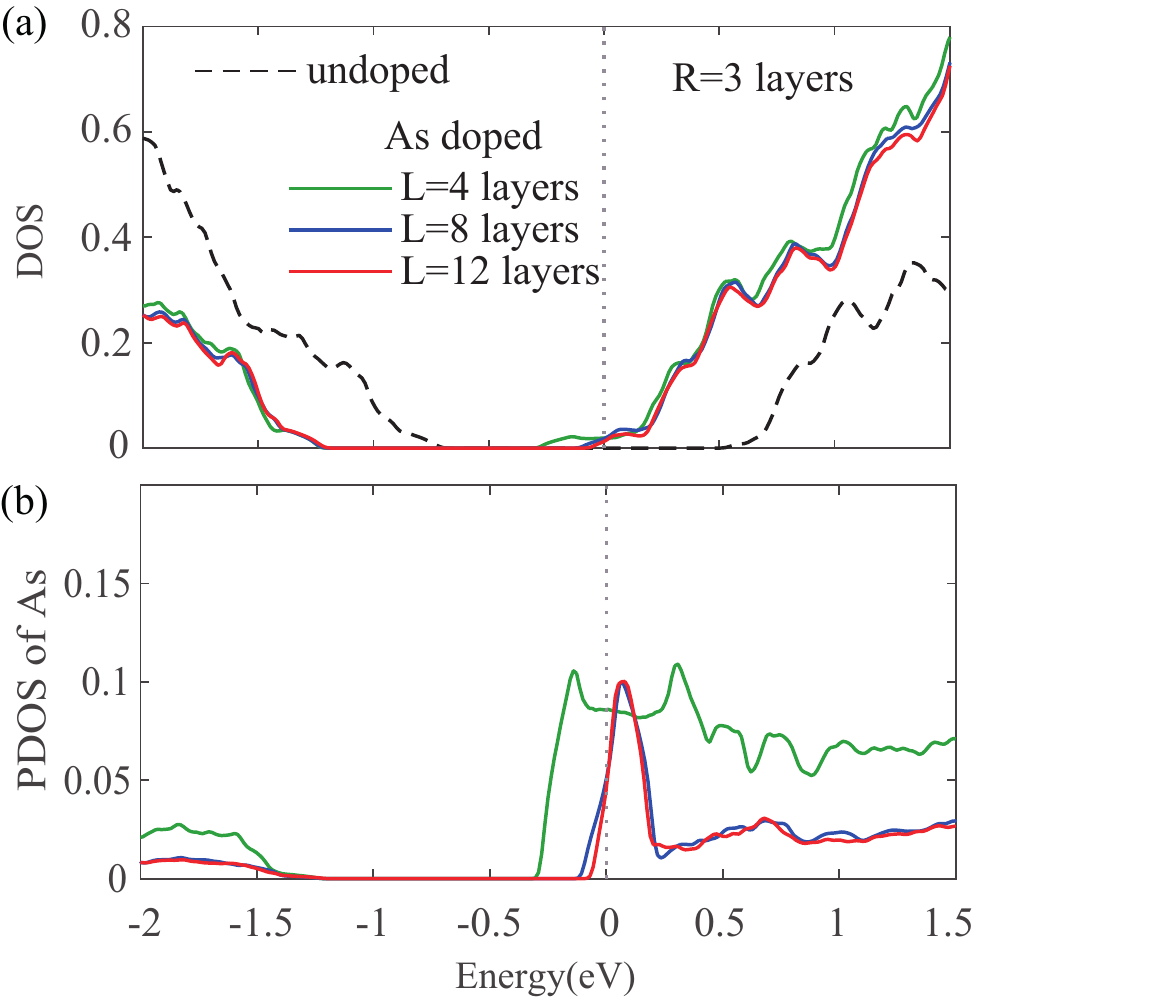}
\caption{\label{fig:DOSLength} (a) Comparison of DOS of doped (solid lines) and
  undoped (dashed line) NWs. Three different longitudinal simulation cell
  lengths (4,8 and 12 atomic layers)
  are compared for doped NWs. (b) Comparison of PDOS of the As atom of
  doped Si NWs. The radii R = 3 layers and the As dopant is placed in
  the center of the NW (shown in Fig.~\ref{fig:PDOSR3L8}(a), As position
  3), in all cases.  In both cases, energies are relative to the Fermi
  level. } 
\end{figure}

As would be expected, the main effect of the introduction of an As
dopant is to create a state near the conduction band edge.  We can
consider the axial length of the NW in terms of dopant
concentration, though dopant spacing along the axis (a one dimensional
doping concentration) is perhaps a more helpful description.  Relatively low levels of
doping (L=8 and 12 layers) introduce a rather flat, and hence low
velocity, band that leads to a high density of states near the
conduction band edge (PDOS for the As atom in the NWs is shown
in Fig.~\ref{fig:DOSLength}). The 
curvature of this band changes with the doping concentration: the
flattest band is with L=12 layers, though there is little difference at
L=8 layers.  At the highest level of doping considered here (L=4
layers), the curvature is significant. Given that this corresponds to
a dopant-dopant spacing of 0.77\,nm it is reasonable that there is considerable
dispersion (in the limit of a strongly delocalised band, we would
expect near-parabolic dispersion).  The almost flat bands seen for L=8 in
Fig.~\ref{fig:bands}(c) and L=12 in Fig.~\ref{fig:bands}(d) suggest that the
dopant-dopant interaction is small at a spacing of 1.55\,nm and
beyond.  This fits well with the donor radius for As in bulk silicon,
calculated as $\sim$1\,nm\cite{Yamamoto:2009pc}; there may be some
compression effects on the dopant state from the finite NW
radius, but our observation of almost flat
bands with a separation of 2.3\,nm indicates that they are likely to
be small.

\begin{figure*}[hbt!]
\begin{minipage}[c]{0.75\textwidth}
\centering
\includegraphics[width=0.98\textwidth]{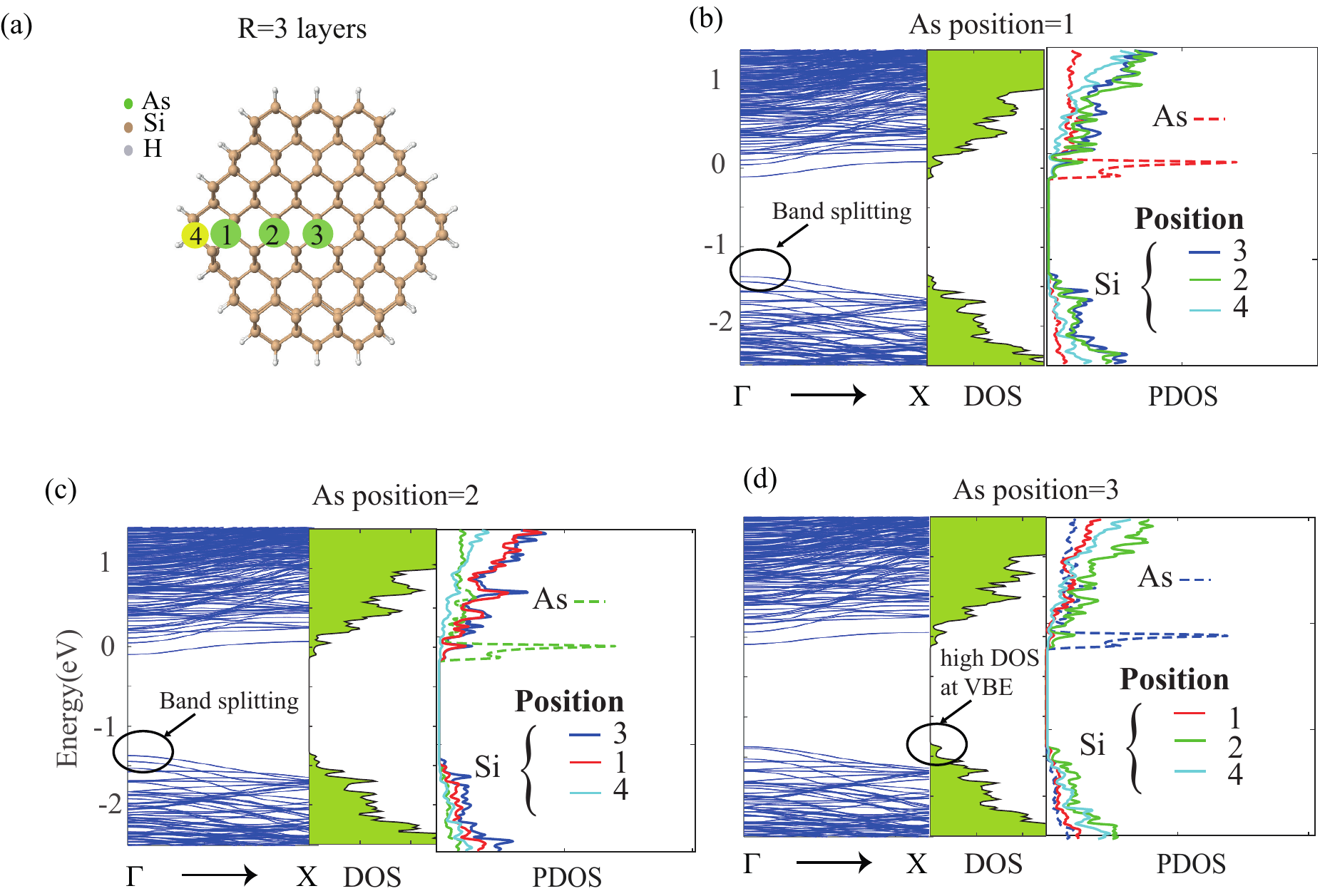}
\end{minipage}\hfill
\begin{minipage}[c]{0.22\textwidth}
\caption{\label{fig:PDOSR3L8} Comparison  of band structure and density of
  states(DOS) of Si NWs with R=3 layers and L=8 layers, with the As
  dopant in different positions.  As positions are indicated in
  (a). The band structure, DOS and the PDOS for positions 1 (close 
  to the surface of the NW), 2 (between the center and the surface of
  the NW) and 3 (close to the center of the NW) are shown in (b),(c)
  and (d), respectively.}  
\end{minipage}
\end{figure*}

\begin{figure}
\centering
\includegraphics[width=0.48\textwidth]{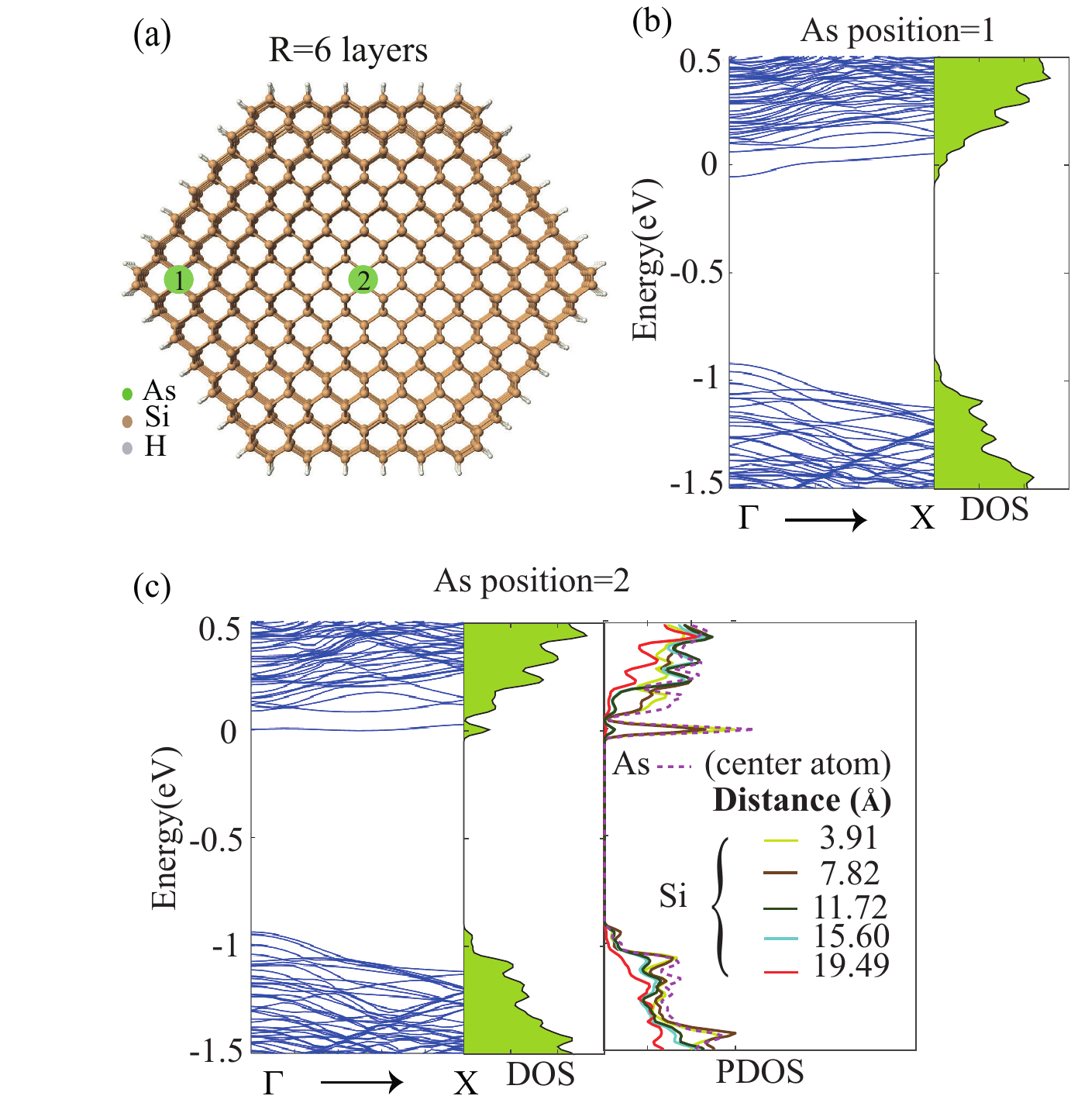}
\caption{\label{fig:PDOSR6L8} Comparison  of band structure and density of
  states(DOS) of Si NWs with R=6 layers and L=8 layers, with the As
  dopant in different positions. As positions are indicated in
  (a). (b) shows  the band structure and density of states(DOS) when
  the As dopant is close to the surface(position 1). (c) shows the
  bandstructure, DOS and  PDOS of atoms at different distances from
  the dopant when the As dopant in the center of the NW
  (position 2). } 
\end{figure}

\begin{figure}
\centering
\includegraphics[width=0.45\textwidth]{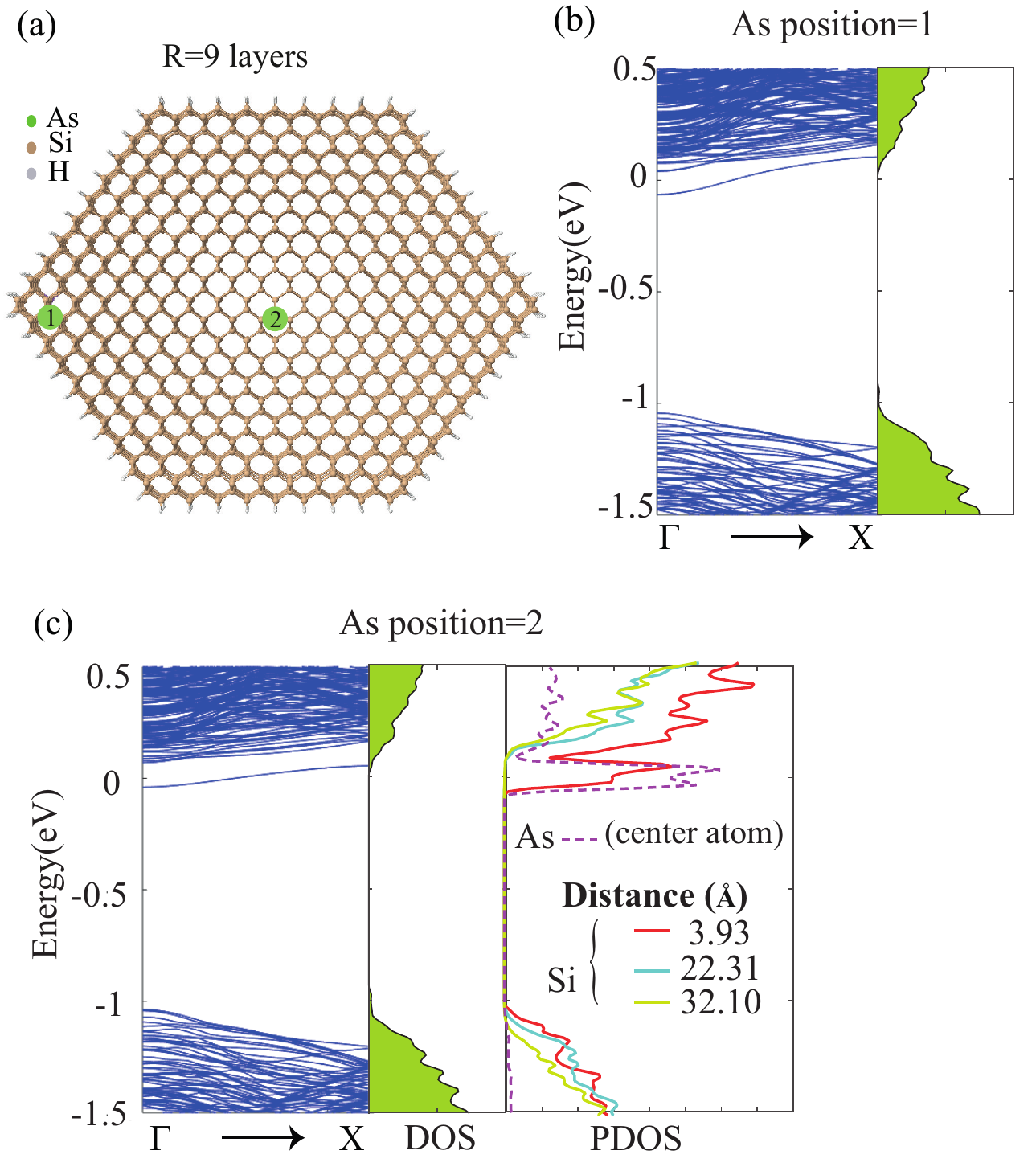}
\caption{\label{fig:PDOSR9L8} Comparison  of band structure and density of
  states(DOS) of Si NWs with R=9 layers and L=8 layers, with the As
  dopant in different positions. As positions are indicated in
  (a). (b) shows  the band structure and density of states(DOS) when
  the As dopant is close to the surface(position 1). (c) shows the
  bandstructure, DOS and  PDOS of atoms at different distances from
  the dopant when the As dopant in the center of the NW
  (position 2).} 
\end{figure}

Figure ~\ref{fig:DOSLength}(a) compares the total DOS of NWs
with different (one dimensional) As doping concentrations.  Since As
is a n-type dopant, we would expect to observe a defect state near the
conduction band, with an accompanying shift in the Fermi level, and
this is indeed seen. In addition to this, the defect/dopant band
introduced within the bandgap results in increased DOS near the
conduction band edge.

For these NWs, the projected DOS of the dopant As atoms is shown
in Figure ~\ref{fig:DOSLength}(b). The results from two and three
repeat lengths are in good agreement with each other. This confirms
that at least two repeat lengths in the axial direction are needed to
reduce artificial interaction between images of the dopant atoms when
simulating Si NWs using DFT methods (we note that previous work
modelling boron and phosphorus-doped
NWs\cite{fernandez2006surface,leao2008confinement} was confined
to much smaller lengths than are used here, mostly L=2 or L=4, though
one very small diameter wire with L=10 was
modelled\cite{leao2008confinement}). Accordingly, for the calculations shown in
Figs.~\ref{fig:PDOSR3L8}, \ref{fig:PDOSR6L8} and \ref{fig:PDOSR9L8} we
used supercells made of two repeat lengths in the axial direction, i.e
L=8, in order to reduce dopant image interaction.

Figure~\ref{fig:PDOSR3L8} shows the effect of dopant positioning on
the density of states, PDOS and the bandstructure, for a NW of
R=3 layers. The As atom is placed to substitute a Si atom near the
surface (1), in the center (3), or mid-way in between (2).  In
Fig.~\ref{fig:PDOSR3L8}(b) and (c), it is clear that the
valence band edge at the $\Gamma$ point is split, due to the
asymmetry when As is placed in a position other than the center.  When
the dopant is at the centre of the NW, the band is not split, leading
to a higher DOS near the valence band edge. It is also apparent from PDOS that
the flat band seen in the bandstructure near the conduction band edge
results from the As dopant. 

Figure~\ref{fig:PDOSR6L8} shows similar results as the radius of the
NW is increased to R=6
layers. The low velocity dopant band introduced due to doping is
slightly further below the conduction band edge than for the R=3 case,
and is more significantly affected by the dopant positioning.  The
centre of the NW is better isolated from the surface in this
case, so that the difference to the near-surface location is enhanced.
The dopant band becomes flatter and creates strongly localised states near the
conduction band edge when the dopant is in the center or close to the
center. The PDOS of atoms can be seen for the central dopant placement
in Fig.~\ref{fig:PDOSR6L8}(c). The major contributions to the dopant
state come from the dopant atom and atoms close to it. Similar to
the R=3 layers case, placing the dopant away from the center breaks
the symmetry and cause band splitting near the valence band edge at
the $\Gamma$ point. 

In Fig.~\ref{fig:PDOSR9L8} we explore our largest doped NW, with
R=9 layers. The super cell containing two unit cells in the axial
direction is made up of 2,472 atoms (2,239 Si atoms,1 As atom and 232 H
atoms). Similar to the smaller diameter NWs above, a
low velocity band is introduce by the doping, and the valence band edge
band splitting can be seen when the dopant is placed in an asymmetric
position. The overall DOS in Fig.~\ref{fig:PDOSR9L8}(b) is much
smoother, and the dopant state does not appear very prominent, due to
the large number of atoms in the system.  The PDOS in
Fig.~\ref{fig:PDOSR9L8}(c) still show the dopant state clearly, with
contributions from neighbouring silicon atoms.

\section{Conclusion}
In conclusion, we have used the \textsc{Conquest} local orbital DFT
code to study both pure and doped Si NWs with diameters up to 9.53~nm and 7.34~nm
respectively.  For the undoped NWs, the bandgap reduces with
increasing diameter and converges to its bulk value, which is in
agreement with experimental data. The atoms closest to the surface of
the NW contributes less to the states near the band edges, when
compared with atoms close to the centre.  This is due
to differences in Si-Si atomic distances and surface passivation
effects. 

When considering As-doped Si NWs, the dopant placement is
important, because it effects electronic structure and hence 
important properties such as electronic transport.
A low velocity band in the gap is introduced by As doping, close to
the conduction band edge. The dopant positioning  affects the
curvature of this band. The curvature reduces when the dopant is placed
closer to the center, creating strong peaks in the density of states.
Asymmetric placement of the dopant creates band splitting in the valence band
edge which is visible quite prominently in the DOS of low radii NWs.
This study demonstrates the importance and utility of large-scale
density functional approaches such as those used in
\textsc{Conquest}.  It is important to note that significantly larger
calculations are possible using multi-site support functions with
exact diagonalisation (up to 10,000 atoms), while linear scaling
calculations are feasible on 100,000+ atoms, with techniques available
to recover electronic structure near the Fermi
level\cite{Nakata:2017mw}.  These approaches show great promise in the
study of experimentally relevant NWs.

\acknowledgements
The authors are grateful for computational support from the UK
Materials and Molecular Modelling Hub, which is partially funded by
EPSRC (EP/P020194), for which access was obtained via the UKCP
consortium and funded by EPSRC grant ref EP/P022561/1.

\bibliographystyle{apsrev4-1}
\bibliography{sample}

\end{document}